\documentclass[12pt]{article}

\makeatletter
\def\cl@chapter{\@elt {theorem}}
\renewcommand\section{\@startsection{section}{1}{\z@}%
                                  {-3.5ex \@plus -1ex \@minus -.2ex}%
                                  {2.3ex \@plus.2ex}%
                                  {\normalfont\large\bfseries}}
\makeatother

\usepackage[backend=bibtex8,sorting=none]{biblatex}
\renewbibmacro{in:}{}
\bibliography{SSH} 

\AtEveryBibitem{\clearfield{month}}
\AtEveryBibitem{\clearfield{doi}}
\AtEveryBibitem{\clearfield{url}}
\AtEveryBibitem{\clearfield{issn}}
\AtEveryBibitem{\clearfield{number}}
\AtEveryBibitem{\clearfield{eid}}

\usepackage{graphicx} 
\graphicspath{ {./images/} }
\usepackage[a4paper,top=1.2in, left=0.8in, right=0.8in, bottom=1in]{geometry}
\usepackage[T1]{fontenc}
\usepackage{tikz}
\usetikzlibrary{intersections,positioning,shapes.misc,calc,arrows}
\usepackage{mathrsfs}
\usepackage{mathtools}
\usepackage{hyperref}
\usepackage{caption}

\hypersetup{colorlinks=true,linktocpage=true,citecolor=blue,urlcolor=cyan,linkcolor=red}
\usepackage{amsfonts}
\usepackage{microtype}
\usepackage[mathscr]{eucal}
\usepackage[affil-it]{authblk}

\newcommand{\real}{\text{{Re}}\:}
\newcommand{\imag}{\text{{Im}}\:}

\newcommand{\first}{\mathcal{H}^{(F)}}
\newcommand{\second}{\mathcal{H}^{(S)}}

\begin{document}

\title{\textbf{\Large{Hilbert space inner products for PT-symmetric Su-Schrieffer-Heeger models}}}
\author{Frantisek Ruzicka
  \thanks{Electronic address: \texttt{ruzicfra@fjfi.cvut.cz}}}
\affil{\normalsize{Faculty of nuclear sciences and physical engineering \\
Czech technical university in Prague \\ Brehova 7, 115 19 Prague \\ Czech Republic}}
\affil{Nuclear physics institute, Czech academy of sciences \\ Hlavni 130, 250 86 Husinec-Rez \\  Czech republic }
\date{\normalsize{\today}}

\maketitle

\begin{abstract}
A Su-Schrieffer-Heeger model with added PT-symmetric boundary term is studied in the framework of pseudo-hermitian quantum mechanics. For two carefully chosen special cases, a complete set of pseudometrics is constructed in closed form. When complemented with a condition of positivity, the pseudometrics determine all the physical inner products of the considered model.
\end{abstract}

\section{Introduction}

The Su-Schrieffer-Heeger (SSH) model \cite{SSH1,SSHeasy} is one of the benchmark topologically nontrivial models in physics of condensed matter. It provides a convenient description of certain physical systems \cite{ZakPhase,ultracoldFermions}, and serves as one of the simplest examples of topological insulators \cite{topInsulators}. The $n$-site SSH model may be expressed as

\begin{equation}
\label{SSH}
H^{(n)}_{SSH} = \sum_{i=1}^N \left[   t(1-\Delta \cos \theta) a^\dagger_{2i-1} a^{\phantom{\dagger}}_{2i} 
+ t(1+\Delta \cos \theta) a^\dagger_{2i} a^{\phantom{\dagger}}_{2 i+1} + h.c. \right]
\end{equation}

with $a^{\dagger}_{i}, a^{\phantom{\dagger}}_{i}$ being the $i$-th site fermionic creation and annihilation operators. It has been suggested \cite{SSHeasy} to complement this hermitian SSH model with a non-hermitian $\mathscr{PT}$-symmetric (invariant under simultaneous action of parity and time reversal) boundary term

\begin{equation}
\label{SSHnonhermitian}
H_I = \gamma  a^\dagger_1 a^{\phantom{\dagger}}_1 + \gamma^* a^\dagger_n a^{\phantom{\dagger}}_n
\end{equation}

with $\gamma = i \alpha, \alpha \in \mathbb{R}$. Despite apparent non-hermicity of the resulting operator $H = H_{SSH} + H_{I}$, it has been shown that its eigenvalues remain real for $|\alpha|$ sufficiently small. This motivates further examination of this model from the viewpoint of $\mathscr{PT}$-symmetric quantum mechanics.

$\mathscr{PT}$-symmetric quantum mechanics \cite{BenderBoettcher,BenderReview,DoreyDunningTateo} is a theoretical framework for studying representations of quantum observables by non-hermitian operators. It is based on a a simple fact, that an observable may be represented by a non-hermitian operator on a given Hilbert space $\first = (\mathcal{V}, \langle \cdot | \cdot \rangle)$, as long as such operator is hermitian in another Hilbert space $\second = (\mathcal{V}, \langle \phi | \psi \rangle_{\Theta})$ \cite{threeHilbertSpace}. The inner products on $\mathcal{V}$ are in one-to-one correspondence with so-called metric operators $\Theta$ through the formula

\begin{equation}
\label{innerProduct}
\langle \phi | \psi \rangle_{\Theta}  = \langle \phi | \Theta | \psi \rangle
\end{equation}

where we require $\Theta$ to be bounded, hermitian and positive in order to generate a genuine inner product \cite{ScholzGeyerHahne}. The hermicity condition of $H$ in $\second$ could be expressed in operator form as
\begin{equation}
\label{DieudonneEquation}
H^\dagger \Theta = \Theta H
\end{equation}

which is a direct generalization of the standard condition $H = H^\dagger$. Operators $H$ satisfying \autoref{DieudonneEquation} for at least one metric operator $\Theta$ are called quasi-hermitian \cite{quasiHermitian}. It is clear that on a given Hilbert space $\first$, all admissible observables are described by quasi-hermitian instead of hermitian operators. Furthermore, recall that for any bounded, positive $\Theta$, there always exists a decomposition $\Theta = \Omega^\dagger \Omega$ with $\Omega$ bounded. Inserting this decomposition into \autoref{DieudonneEquation} yields

\begin{equation}
h := \Omega H \Omega^{-1} = (\Omega H \Omega^{-1})^\dagger = h^\dagger
\end{equation}

which shows that being quasi-hermitian is equivalent to being boundedly diagonalizable with real spectrum (instead of unitarily diagonalizable). However, the construction of the similarity transformation $\Omega$ may be in general very difficult for a given $H$. Special attention is usually given to parametric quasi-hermitian models with nontrivial domains of observability. The boundaries of such domains are formed of the so-called exceptional points (EPs) \cite{Kato,HeissEP}, points in parameter space, where the operator ceases to be diagonalizable. The boundary crossings of physical parameters (e.g. time) are studied under the name quantum catastrophes \cite{ZnojilCatastrophes}, which emphasizes close relationship to classical catastrophe theory \cite{ArnoldCatastrophes}.

This paper is divided as follows: in section 2, we introduce the general case of the examined $\mathscr{PT}$-symmetric SSH model. In section 3, we link its special case $\cos \theta = 0$ to a discretized Robin square well, and construct a complete set of pseudometrics. In section 4, we repeat this procedure for a dual SSH nearest-neighbor interaction. In section 5, we consider the general case $H = H_{SSH} + H_{I}$, and address the questions of pseudometric construction and  cutoff emergence. Section 6 is devoted to discussion and remarks.

\section{The $\mathscr{PT}$-symmetric SSH model}

Inspired by \cite{SSHeasy}, we examine a SSH model with open boundary conditions, complemented by the $\mathscr{PT}$-symmetric boundary term as in \autoref{SSHnonhermitian}. Throughout this paper, we set with no loss of generality $t=1, \: \Delta=1$, and denote $\lambda = \cos \theta$ in \autoref{SSH}.  The coupling constant $\gamma$ was taken to be purely imaginary in \cite{SSHeasy}, while we assume a more general case $\gamma = \rho + \mathrm{i} \omega \in \mathbb{C}$. In order to apply powerful tools of linear algebra, we work in a matrix representation of the creation and annihilation operators. The basis of a corresponding two-dimensional Hilbert space is chosen, such that

\begin{align}
\begin{split}
&a | 0 \rangle  = 0, \qquad \;\; a^\dagger | 0 \rangle = | 1 \rangle \\
&a | 1 \rangle  = | 0 \rangle, \qquad a^\dagger | 1 \rangle = 0
\end{split}
\end{align}

In such case, the model from \autoref{SSH} becomes a family of $n \times n$ matrices with $n= 2N$. We adopt a strategy of printing the matrices explicitly for the $n=4$, as long as the extrapolation pattern for higher $n \in \mathbb{N}$ is clear enough. That means we may write

\begin{equation}
\label{SSHnormal}
H^{(4)} = H^{(4)}_{SSH} + H_I^{(4)} = \begin{bmatrix}
\gamma & -1-\lambda & & \\
-1-\lambda & & -1+\lambda & \\
 & -1+\lambda & & -1-\lambda \\
 & & -1-\lambda & \gamma^*
\end{bmatrix}
\end{equation}

\begin{figure}[h!]
\centering
    \includegraphics[height=2.4cm,width=5.2cm]{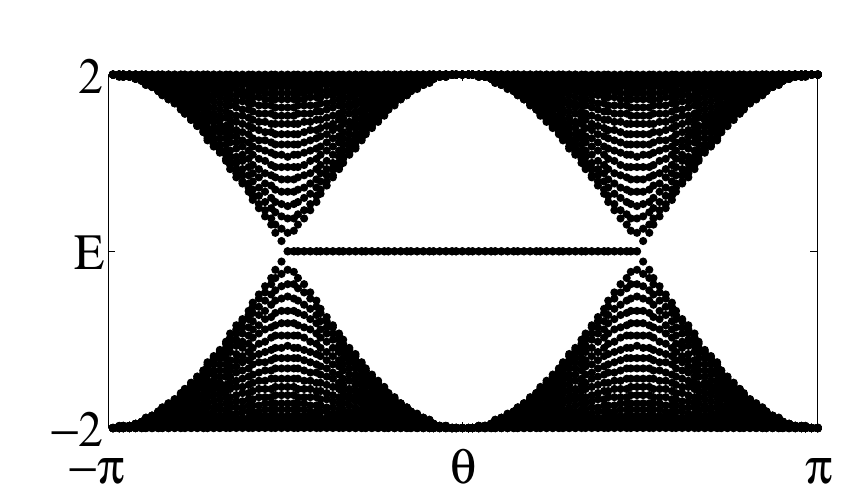} 
    \includegraphics[height=2.4cm,width=5.2cm]{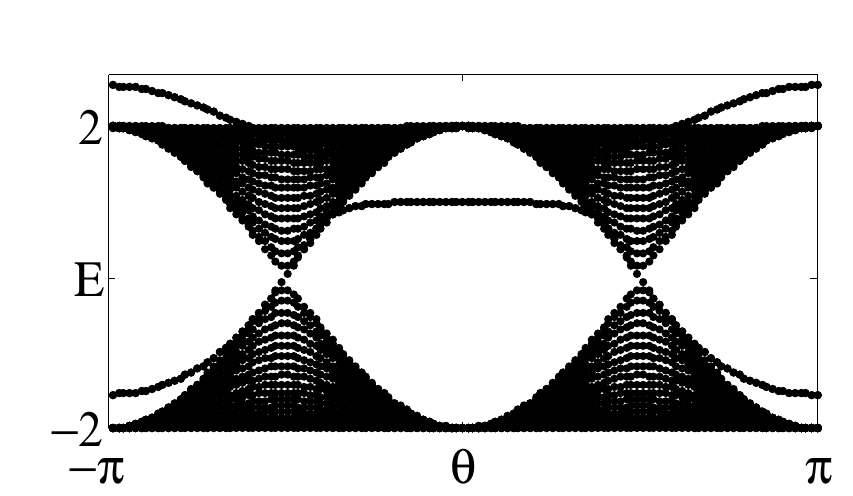}  
    \includegraphics[height=2.4cm,width=5.2cm]{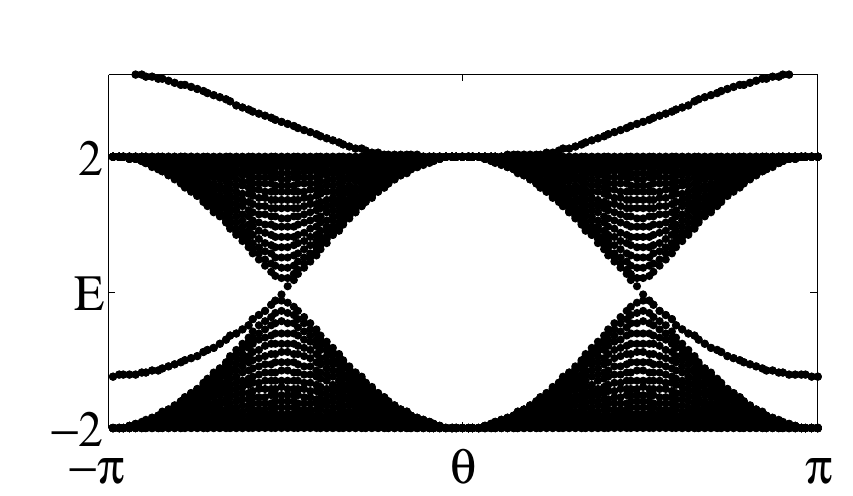} \\
    \includegraphics[height=2.4cm,width=5.2cm]{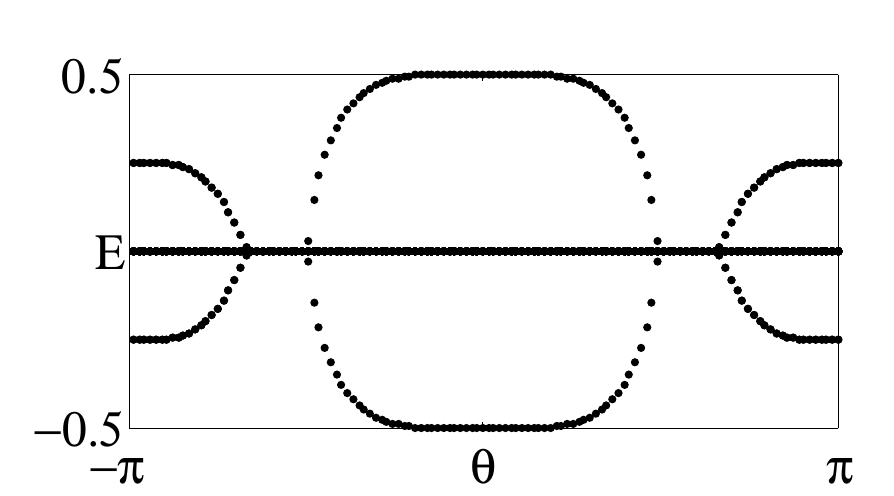} 
    \includegraphics[height=2.4cm,width=5.2cm]{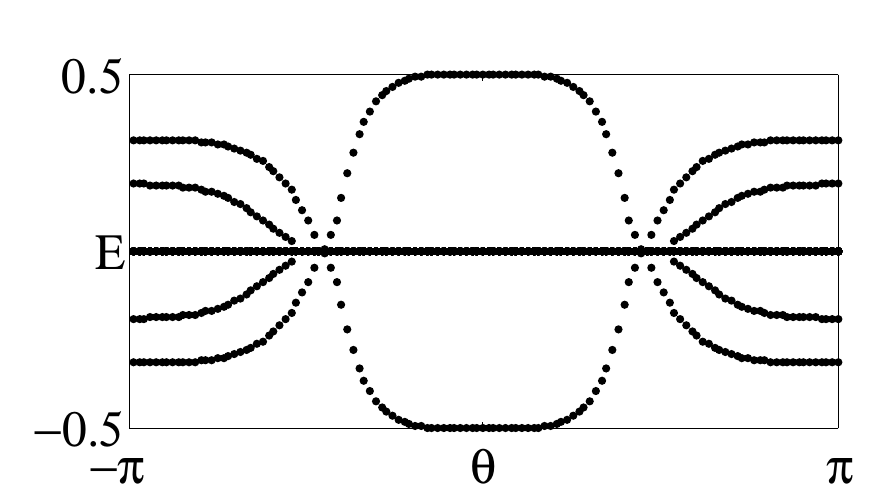}
    \includegraphics[height=2.4cm,width=5.2cm]{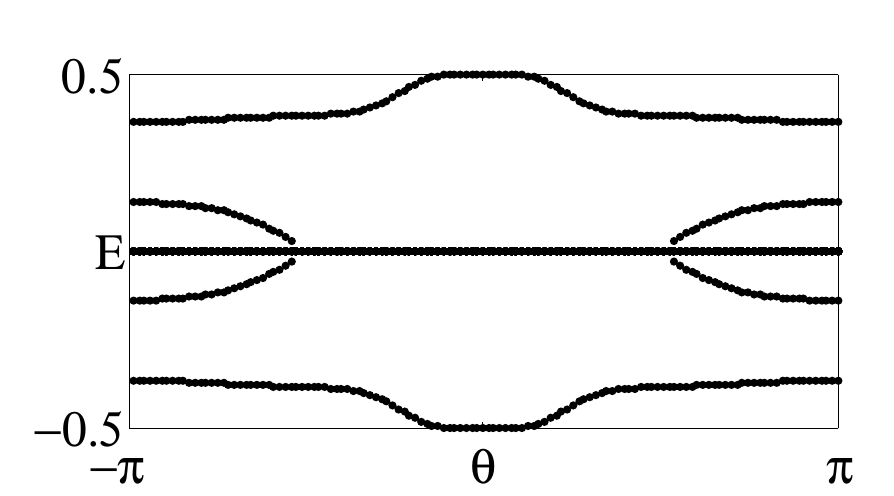} \\
    \caption{\emph{$\real [\sigma (H)]$ and $\imag [\sigma (H)]$ for the $n=50$ SSH model as a function of $\theta$. The values of $\gamma$ are $0.5 i, 0.5 i + 1$ and $0.5i + 2$.}}
    \label{SSHfig}
\end{figure}

In parallel with numerical experiments of \cite{SSHeasy}, we illustrate the effects of general complex coupling in \autoref{SSHfig}. Clearly, there is no reason to confine attention to purely imaginary $\gamma$ in our $\mathscr{PT}$-symmetric considerations, as for $\rho$ reasonably small, there always exists a nonempty interval of $\omega$ with real spectrum. Size of such interval shrinks with growing $\rho$, and finally vanishes completely, in the present case for $\rho \approx 1$.

\begin{figure}[h!]
    \centering
    \includegraphics[height=4.6cm,width=7cm]{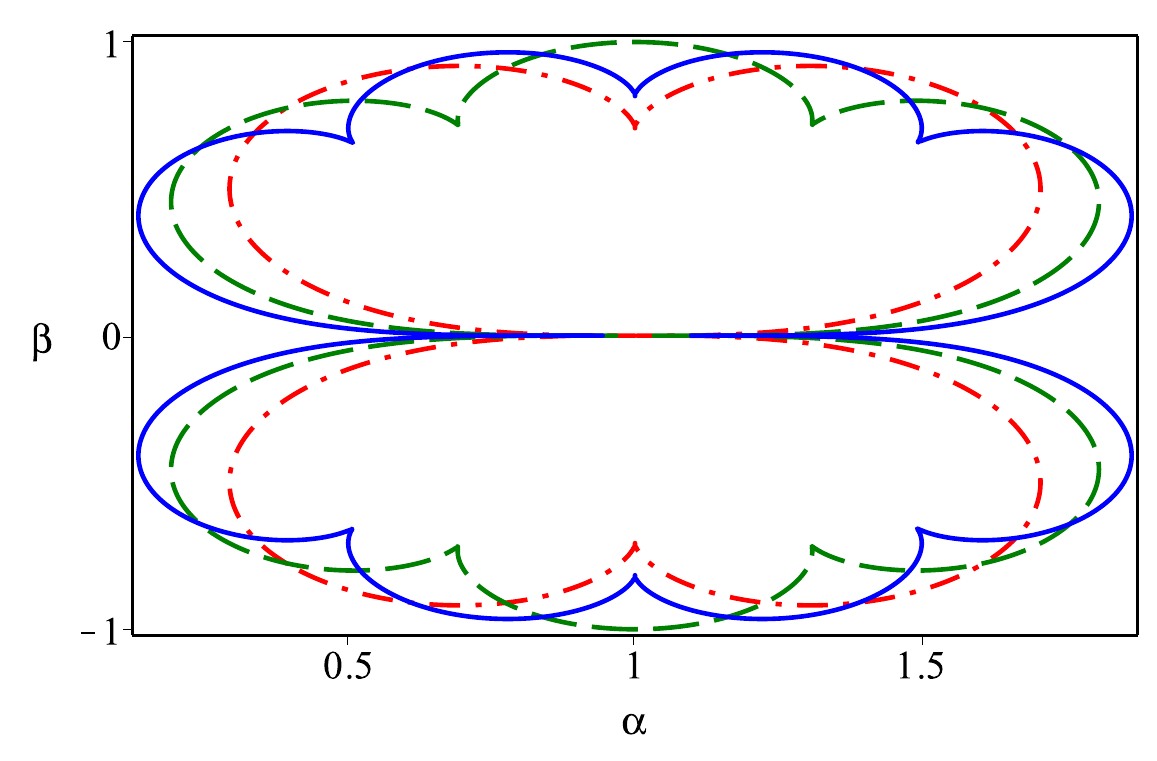} 
    \includegraphics[height=4.6cm,width=7cm]{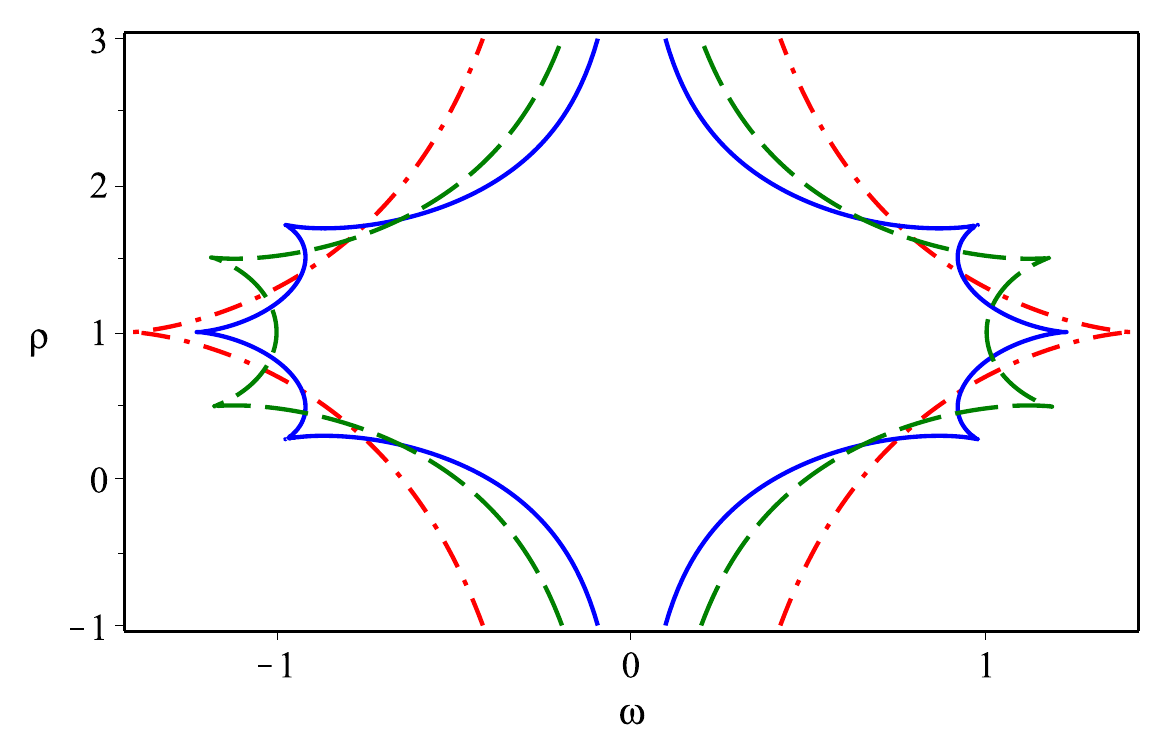} 
    \caption{\emph{Exceptional points of \autoref{discreteKrejcirik} for $n=3,4,5$ in the $(\alpha,\beta)$ and $(\rho, \omega)$ coordinates.}}
    \label{dobs}
\end{figure}

\section{The Robin square well}

The construction of $\Theta$ for infinite-dimensional quasi-hermitian models is in general a highly nontrivial task, which has to be approached perturbatively. An exceptional operator admitting exact metric construction (Laplacian on a real interval with imaginary Robin boundary conditions) was introduced in \cite{KrejcirikBila}. This inspired a subsequent paper \cite{discreteRobin}, which applied discretization techniques on this continuous model, resulting in a $\mathscr{PT}$-symmetric family of matrices

\begin{equation}
\label{discreteKrejcirik}
H_{DR}^{(4)} = \begin{bmatrix}
\gamma & -1 & & \\
-1 & & -1 & \\
& -1 &  & -1 \\
&  & -1 & \gamma^*
\end{bmatrix} \qquad \text{with } \;\;\;
\gamma = \frac{1}{1 - \alpha - i \beta}
\end{equation}

which clearly coincides with \autoref{SSHnormal} for $\lambda=0$. Throughout this paper, we shall use the natural parametrization $\gamma = \rho  +  \mathrm{i} \omega$ instead of \autoref{discreteKrejcirik}. The coordinate transformation connecting these parametrizations is

\begin{equation}
\omega = \frac{\beta}{(1-\alpha)^2 + \beta^2}, \qquad
\rho = \frac{1-\alpha}{(1-\alpha)^2 + \beta^2}
\end{equation}

The domains of quasi-hermicity in both coordinate systems are shown in \autoref{dobs}. The non-shrinking behavior of the domains in the limit $n \rightarrow \infty$ agrees with the existence of a quasi-hermitian operator in the continuous limit. Now, recall that any $n \times n$ metric may be expressed as $\Theta = \sum_{k=1}^n  \kappa_n  | n \rangle \langle n |$ \cite{metricZnojil}, where $| n \rangle$ are the eigenvectors of $H^\dagger$. Thus, a general metric depends on $n$ arbitrary parameters. Equivalently, we can construct $n$ linearly independent hermitian (but not necessarily positive) matrices $\mathcal{P}^{(k)}$ for a given $H$, such that

\begin{equation}
\label{metricAnsatz}
\Theta = \sum_{k=1}^{n} \varepsilon_k \mathcal{P}^{(k)}
\end{equation}

with $\varepsilon_k$ being constrained by the condition $\Theta > 0$. Such a set of $\mathcal{P}^{(k)}$ is called a complete set of pseudometrics. Although the decomposition from \autoref{metricAnsatz} may be in general quite arbitrary, our aim is to find the pseudometrics in a form, which admits extrapolation to general $n \in \mathbb{N}$. Already in \cite{discreteRobin}, the authors appreciated the existence of a particular metric $\Theta$ with elements

\begin{equation}
\label{ZnojilMetric}
\Theta_{jk} =
  \begin{cases}
   1 &  j = k \\
   - \mathrm{i} \omega (\rho - i \omega)^{j-k-1}       &  j > k \\
   \phantom{-} \mathrm{i} \omega (\rho + i \omega)^{j-k-1} &  j < k 
  \end{cases}
\end{equation}

which served also as a starting point of our considerations. For sufficiently low dimensions, we can construct pseudometrics explicitly with the help of symbolic manipulation on any computer algebra system. For the present two-site model, after denoting $\xi = \rho - \mathrm{i} \omega$, we get

\begin{align}
\begin{split}
\label{metricSpecial}
&\mathcal{P}^{(1)} = 
\begin{bmatrix}
1 & -\mathrm{i} \omega  & - \mathrm{i} \omega \xi & - \mathrm{i} \omega \xi^2 \\
\mathrm{i} \omega & 1  &  - \mathrm{i} \omega & - \mathrm{i} \omega \xi \\
\mathrm{i} \omega \xi^* &  \mathrm{i} \omega & 1 &  - \mathrm{i} \omega \\
\mathrm{i} \omega \xi^{*2} & \mathrm{i} \omega \xi^* &  \mathrm{i} \omega & 1 \end{bmatrix}
\;\;\;\; \mathcal{P}^{(2)} = \begin{bmatrix}
 & 1  & -\mathrm{i} \omega & - \mathrm{i} \omega \xi \\
1 & \rho  & 1 & - \mathrm{i} \omega \\
\mathrm{i} \omega & 1 & \rho & 1 \\
\mathrm{i} \omega \xi^* & \mathrm{i} \omega & 1 &  \end{bmatrix} \\
& \qquad \qquad \;\;\;\; \mathcal{P}^{(3)} = \begin{bmatrix}
&& 1 & - \mathrm{i} \omega \\
& 1 & \rho & 1 \\
1 & \rho & 1 &  \\
\mathrm{i} \omega & 1 && \end{bmatrix}  
\;\;\;\;
\mathcal{P}^{(4)} = \begin{bmatrix}
&&& 1 \\
&&1& \\
&1&& \\
1&&& \end{bmatrix}
\end{split}
\end{align}

Note that $\mathcal{P}^{(1)}$ coincides with \autoref{ZnojilMetric}, while $\mathcal{P}^{(4)}$ realizes a discrete operator of parity, and the condition $H^\dagger \mathcal{P}^{(4)} = \mathcal{P}^{(4)} H$ demonstrates the $\mathscr{PT}$-symmetry of the present model. The two-site case suggests an extrapolation pattern for higher $n \in \mathbb{N}$ with the $k$-th pseudometric having only $2(n-k) +1$ nonzero antidiagonals. The nonzero elements of such metrics are given by the following table

\begin{equation}
\begin{tikzpicture}[baseline=(current  bounding  box.center)]
\node (table) [inner sep=1pt] {
\bgroup
\def\arraystretch{1.2}
\begin{tabular}{c|l}
  $- \mathrm{i} \omega  \xi^{(i-j-k)}$ & $ i-j \geq k$ \\
  \hline
  $ \phantom{-} \mathrm{i} \omega  \xi^{*(i-j-k)}$ & $ j-i \geq k$ \\
  \hline
  $\rho$ & $|i-j|<k,\:  i+j-k$ \text{ even} \\
  \hline
  $1$ & $|i-j|<k,\: i+j-k$ \text{ odd} \\
\end{tabular}
\egroup
};
\draw [rounded corners=.5em] (table.north west) rectangle (table.south east);
\end{tikzpicture}
\label{metricGeneral}
\end{equation}

A rigorous proof of the above proposition by double induction $n \rightarrow n+1$ and $k \rightarrow k+1$ would be a straightforward, although lengthy, generalization of proposition 2 in \cite{discreteRobin} (which is essentially the proof of $n \rightarrow n+1$ for $k=1$), and therefore may be omitted. The positivity of the resulting metric as a function of $\varepsilon_i$ is a generally nontrivial problem, which lies outside of the scope of the present discussion. We can however make use of the fact, that the pseudometric $\mathcal{P}^{(1)}$ is a genuine metric, and treat other $\varepsilon_i$ using tools of perturbation theory.

\section{The dual-SSH model}

A large subclass of tridiagonal quasi-hermitian matrices has a close connection to the theory of orthogonal polynomials \cite{myCatastrophes,JundeWells}. Inspired by these models, we define the dual-SSH (or dSSH) model, which, in parallel with the operator-matrix correspondence of \autoref{SSHnormal} has its two-site form

\begin{equation}
\label{orthogonalPolynomials}
H^{(4)}_{dSSH}  = \begin{bmatrix}
\gamma & -1-\lambda && \\
-1+\lambda && -1-\lambda & \\
& -1+\lambda && -1-\lambda \\
&& -1+\lambda & \gamma^*
\end{bmatrix}
\end{equation}

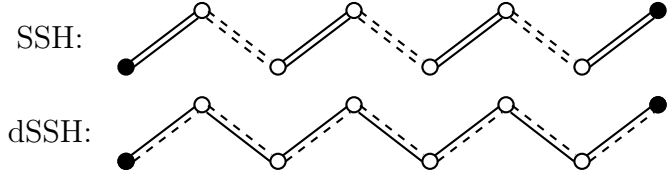
\begin{figure}[h!]
\centering
  \begin{tikzpicture}[scale=.5]
    \node[draw=none] at (-2,0.8) {SSH:};
    \draw[thick,fill=black] (0,0) circle (.2cm);
    \draw[thick,fill=black] (14,1.5) circle (.2cm);
    \foreach \x in {1,...,3}
    \draw[xshift=2*\x cm,thick] (2*\x cm,0) circle (.2cm);
    \foreach \x in {0.5,...,2.5}
    \draw[xshift=2*\x cm,thick] (2*\x cm,1.5) circle (.2cm);
    \foreach \y in {0.03,...,3.03}
    \draw[xshift=2*\y cm,thick] (2*\y cm,0.15) -- +(1.7 cm,1.3);
    \foreach \y in {0.05,...,3.05}
    \draw[xshift=2*\y cm,thick] (2*\y cm,0.0) -- +(1.7 cm,1.3);
    \foreach \y in {0.53,...,2.53}
    \draw[xshift=2*\y cm,thick,dashed] (2*\y cm,1.3) -- +(1.67 cm,-1.3);
    \foreach \y in {0.54,...,2.54}
    \draw[xshift=2*\y cm,thick,dashed] (2*\y cm,1.48) -- +(1.67 cm,-1.3);
  \begin{scope}[yshift=-2.5cm]
    \node[draw=none] at (-2,0.8) {dSSH:};
    \draw[thick,fill=black] (0,0) circle (.2cm);
    \draw[thick,color=white] (0,-0.5) circle (.2cm);
    \draw[thick,fill=black] (14,1.5) circle (.2cm);
    \foreach \x in {0,...,3}
    \draw[xshift=2*\x cm,thick] (2*\x cm,0) circle (.2cm);
    \foreach \x in {0.5,...,3.5}
    \draw[xshift=2*\x cm,thick] (2*\x cm,1.5) circle (.2cm);
    \foreach \y in {0.03,...,3.03}
    \draw[xshift=2*\y cm,thick] (2*\y cm,0.15) -- +(1.7 cm,1.3);
    \foreach \y in {0.05,...,3.05}
    \draw[xshift=2*\y cm,thick,dashed] (2*\y cm,0.0) -- +(1.7 cm,1.3);
    \foreach \y in {0.53,...,2.53}
    \draw[xshift=2*\y cm,thick] (2*\y cm,1.3) -- +(1.67 cm,-1.3);
    \foreach \y in {0.54,...,2.54}
    \draw[xshift=2*\y cm,thick,dashed] (2*\y cm,1.48) -- +(1.67 cm,-1.3);
  \end{scope}
  \end{tikzpicture}
  \caption{\emph{Schemes of the four-site} SSH \emph{and} dSSH \emph{models.}}
  \label{interactionNew}
\end{figure}

Schematic drawings of the both dSSH and the original SSH interaction are shown for $N=4$ in \autoref{interactionNew}, with black circle denoting the boundary term from \autoref{SSHnonhermitian}, and solid, respectively dashed lines denoting the $-1-\lambda$, respectively $-1+\lambda$ interaction

In this section, we study the special case of \autoref{orthogonalPolynomials} with $\gamma=0$. This matrix family indeed belongs to a class of tridiagonal models admitting elegant treatment using theory of orthogonal polynomials. Recall that any orthogonal polynomial sequence $p_n (x)$ obeys a three-term recurrence relation

\begin{equation}
\label{secularEquation}
x p_{n} (x) =  a_{n (n+1)} p_{n+1} (x) + a_{nn} p_n (x) + a_{n(n-1)} p_{n-1} (x)
\end{equation}

with $a_{n(n+1)} a_{(n+1)n} > 0$ and $a_{nn} \in \mathbb{R}$. The coefficients $a_{ij}$ may be understood as elements of truncated three-diagonal $n \times n$ matrices with characteristic polynomial $p_n (x)$. Consequently, the eigenvalues of such matrices (the roots of $p_n$) are real and distinct, which is a sufficient condition for matrix quasi-hermicity. Moreover \cite{ZnojilJacobi}, a complete set of pseudometrics for such models has a very feasible form, with $\mathcal{P}^{(k)}$ having only $(2k+1)$ nonzero diagonals, and may be constructed by solving the recurrences

\begin{equation}
\label{orthoMetrics}
\sum_{j=0}^{k-1} a^{\phantom{()}}_{(k+j) (k)}  \mathcal{P}^{(k)}_{(k+j)(k+1)}  =  \sum_{j=0}^{k-1} a^{\phantom{()}}_{(k+j) (k+1)} \mathcal{P}^{(k)}_{(k)(k+j)} 
\end{equation}

The matrix elements of \autoref{orthogonalPolynomials} obey the condition $a_{n(n+1)} a_{(n+1)n} > 0$ for any $\theta \neq k \pi$. At the exceptional points $\theta = k \pi$ the operator is not diagonalizable despite the reality of its spectrum, and its quantum-mechanical interpretation is lost. For $\theta \neq k \pi$, we may construct pseudometrics using \autoref{orthoMetrics}, or again with the help of computer-based symbolic manipulations. In either case, we arrive at the two-site formulae

\begin{equation}
\mathcal{P}^{(1)} = \begin{bmatrix}
+  &&& \\
&  - && \\
&& + &  \\
&&& - \end{bmatrix} \qquad
\mathcal{P}^{(2)} = \begin{bmatrix}
 & 1  & & \\
1 &  & 1 &  \\
& 1 &  & 1 \\
&& 1 &  \end{bmatrix} \qquad
\mathcal{P}^{(3)} = \begin{bmatrix}
&& + & \\
& - &  & - \\
+ & & + &  \\
& - & & \end{bmatrix} \qquad
\mathcal{P}^{(4)} = \begin{bmatrix}
&&& 1 \\
&& 1 & \\
& 1 && \\
1 &&& \end{bmatrix}
\end{equation}

where we have denoted $\pm = (1 \pm \lambda)$. These formulae suggest an extrapolation pattern with $k$-th pseudometric having $2k+1$ nonzero diagonals and $2(n-k)+1$ nonzero antidiagonals. Moreover, the nonzero elements are arranged in a chessboard-like pattern, in accordance with \cite{JundeWells}. We may write formula for zero elements as

\begin{equation}
\mathcal{P}^{(k)}_{ij} = 0 \begin{cases}
\;\; \text{for } \; |i+j-k| \text{ odd} \\
\;\; \text{for } \; |i-j| \geq k \text{ or } | i+j-n-1| > n-k
\end{cases}
\end{equation}

Nonzero elements of $\mathcal{P}^{(k)}$ are given by the following table. In the left column, we have listed for comparison also nonzero pseudometric elements for the classical SSH interaction from \autoref{SSHnormal} with $\gamma = 0$. The hermicity of such model is apparent from $\Theta = I$ being among the admissible metrics.

\begin{equation}
\begin{tikzpicture}[baseline=(current  bounding  box.center)]
\node (table) [inner sep=1pt] {
  \bgroup
  \def\arraystretch{1.2}
\begin{tabular}{c|l||c|l}
  \multicolumn{2}{c}{\textbf{SSH}}  & \multicolumn{2}{c}{\textbf{dSSH}}\\
  \hline
  $+$ & $k$ even, $i$ odd & $+$ & $k$ odd, $i$ odd \\
  \hline
  $-$ & $k$ even, $i$ even & $-$ & $k$ odd, $i$ even \\
  \hline
  $1$ & $k$ odd & $1$ & $k$ even \\
\end{tabular}
\egroup
};
\draw [rounded corners=.5em] (table.north west) rectangle (table.south east);
\end{tikzpicture}
\label{tableOrtho}
\end{equation}

\section{The inner products}

After examining two special cases $\gamma = 0$ and $\lambda = 0$, we are ready to address the model of \autoref{SSHnormal} in full generality. In order to print the resulting matrices explicitly, we define $\pm = (1 \pm \lambda) w$, where $w$ appearing in the $(ij)$-th element of the $k$-th pseudometric denotes the $(ij)$-th element of the $k$-th pseudometric of \autoref{metricSpecial}. Explicit construction of metric operators remains feasible for sufficiently low dimensions. For the two-site model, we get the following results (we shall explain shortly why $\mathcal{P}^{(1)}$ is missing from the list)

\begin{align}
\begin{split}
\label{generalTwoSite}
\mathcal{P}^{(2)}_{SSH} &= \begin{bmatrix}
& +-   &  + &  1\\
+-  & -  & -^2 &  + \\
+ & -^2 & - & +-  \\
1 & + & +- & \end{bmatrix} \;\;\;\;\;
\mathcal{P}^{(3)}_{SSH} = \begin{bmatrix}
&& + &  1\\
& - &  1& + \\
+ &  1& - & \\
1 & + &&  \end{bmatrix}  \;\;\;
\mathcal{P}^{(4)}_{SSH} = \begin{bmatrix}
&&& 1 \\
&&1& \\
&1&& \\
1&&& \end{bmatrix} \\ \\
\mathcal{P}^{(2)}_{dSSH} &= \begin{bmatrix}
& -^2   &  - &  1\\
-^2  & -  & +- &  + \\
 - & +- & + & +^2  \\
1 &  + & +^2 &\end{bmatrix} \;\;\;\;\;\:
\mathcal{P}^{(3)}_{dSSH} = \begin{bmatrix}
&& - &  1\\
& - &  1& + \\
- &  1& + & \\
1 & + &&  \end{bmatrix} \;\:
\mathcal{P}^{(4)}_{dSSH} = \begin{bmatrix}
&&& 1 \\
&&1& \\
&1&& \\
1&&& \end{bmatrix}
\end{split}
\end{align}

Extrapolation formulae for general $n \in \mathbb{N}$ are clear from these matrices, again with the $k$-th pseudometric having $2(n-k)+1$ nonzero antidiagonals. However, as long as we wish to impose the same ansatz for pseudometrics $\mathcal{P}^{(k)}, k<n-2$, we discover the emergence of a cutoff value, for which the pattern ceases to be valid. This phenomenon can be seen already in the case of hepta-antidiagonal pseudometrics. We illustrate this behavior on the following three-site models
\begin{equation}
\label{cutoff}
\mathcal{P}^{(3)}_{SSH} = \begin{bmatrix}
&&  +^2 - &  +- & + & 1\\
 & + -^2  &  +- & +^2 - & -^2 & + \\
 +^2- & +- & + -^2  & \Lambda_1 & +^2- & +-\\
+-&  +^2- &  \Lambda_1 & + -^2 & +- & +^2 - \\
+&  -^2 &   +^2- & +- & + -^2 & \\
1&  + &   +- & +^2 - &&  \end{bmatrix} \qquad
\mathcal{P}^{(3)}_{dSSH} = \begin{bmatrix}
&&  -^3 &  -^2 & - & 1\\
 & -^3  &  +- & +^2 - & +- & + \\
-^3 & +- & + -^2  & \Lambda_2 & +^2- & +^2\\
-^2&  +^2- & \Lambda_2 & + -^2 & +- & +^3 \\
-&  +- &   +^2- & +- & +^3 & \\
1&  + &   +^2 & +^3 &&  \end{bmatrix}
\end{equation}

where $+ -^2$ stands for $+(-)^2$. If we insert these matrices into the equation $H^\dagger \mathcal{P}^{(3)} = \mathcal{P}^{(3)} H$, we obtain complex solutions for $\Lambda_i$, which violate the hermicity condition. Thus, in the general case of \autoref{SSHnormal}, we are able to build just three pseudometric families $\mathcal{P}^{(n-2)}, \mathcal{P}^{(n-1)}$ and $\mathcal{P}^{(n)}$ before cutoff emergence. Still, the pattern of matrix elements not approaching infinity as $n \rightarrow \infty$ is strongly conjectured to be preserved, suggesting that the resulting metrics $\Theta$ remain bounded in appropriate continuous limit. Such a continuous analogue is well-known for $\lambda = 0$ \cite{KrejcirikBila}, but the examination of the continuous limit for general $\theta \neq 0$ remains open.

\section{Discussion}

We have successfully constructed a complete set of pseudometrics for the discrete Robin square well, as well as for the hermitian SSH model and its dual counterpart. In the general case of  \autoref{SSH}, we encountered a cutoff preventing the construction of pseudometrics beyond three particular families. While this may sufficient in many scenarios, we emphasize two reasons for the importance of constructing a complete set of pseudometric. As long as the Hamiltonian is not the only dynamical observable, the presence of additional observables $\Lambda_i$ imposes additional constraints on the metric, in the form of equations

\begin{equation}
\Lambda^\dagger_i \Theta = \Theta \Lambda^{}_i
\end{equation}

Furthermore, the physical Hilbert spaces $\second$ generated by different metrics are in general not unitarily equivalent. In particular, for parametric models, some metrics may have larger domains of positivity than others \cite{nonsufficientMetric, myPhaseTransition}. The choice of a correct physical metric is then dictated by other physical principles of, for example, locality.

\begin{figure}[h!]
    \centering
    \includegraphics[height=2.1cm,width=5.2cm]{SSH2R} 
    \includegraphics[height=2.1cm,width=5.2cm]{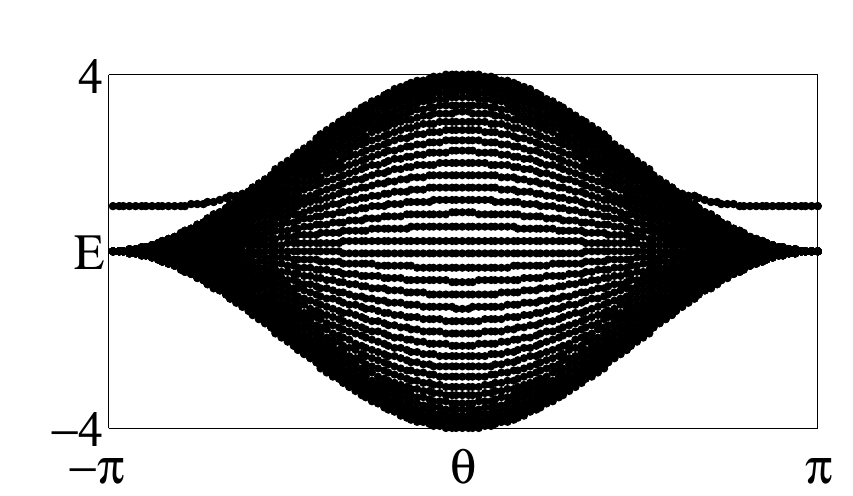} 
    \includegraphics[height=2.1cm,width=5.2cm]{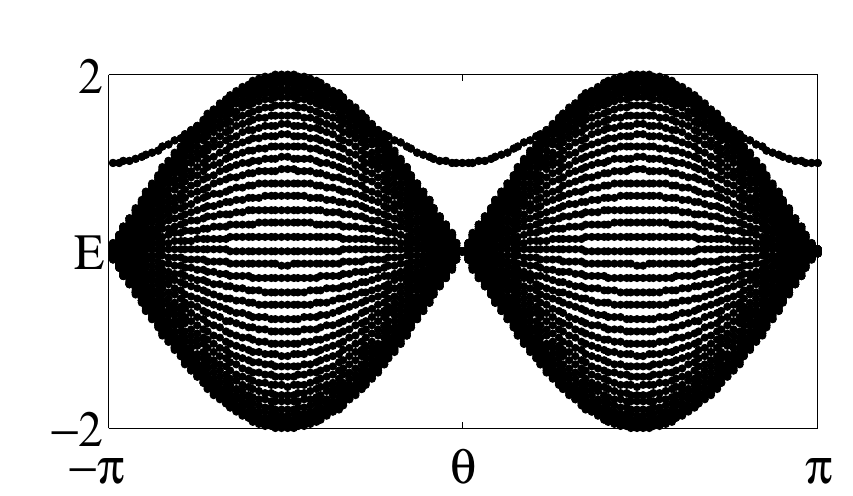} \\
    \includegraphics[height=2.1cm,width=5.2cm]{SSH2I} 
    \includegraphics[height=2.1cm,width=5.2cm]{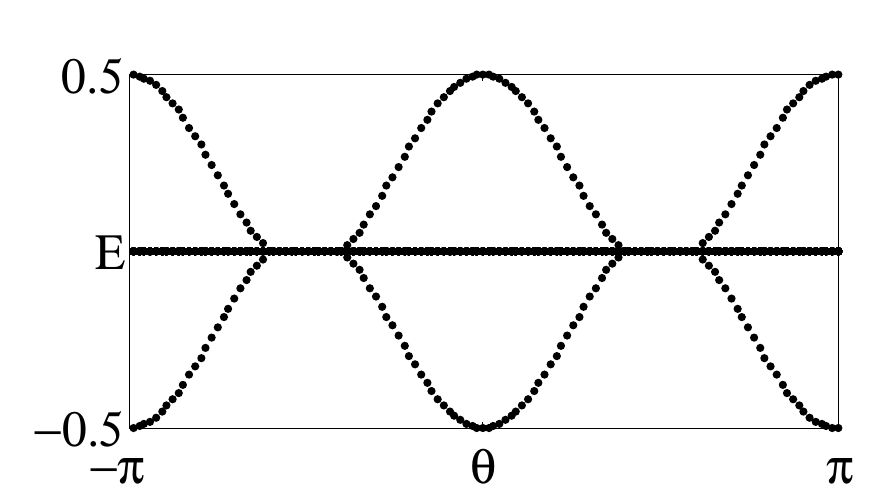} 
    \includegraphics[height=2.1cm,width=5.2cm]{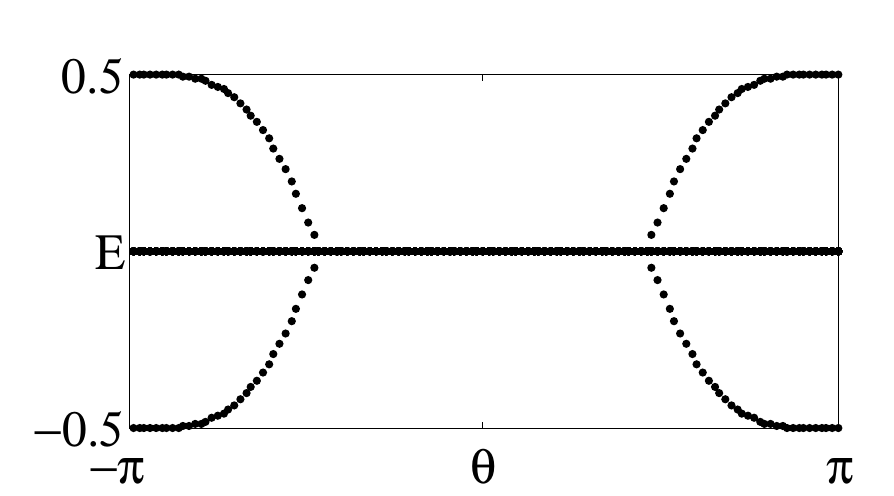} \\
    \caption{\emph{Eigenvalues of the} $n=50$ SSH, dSSH \emph{and the model $-1-\lambda \rightarrow -1 + \lambda$ for $\gamma = 0.5 \mathrm{i} + 0.8$. }}
    \label{SSHfig2}
\end{figure}

Spectral behavior of the discussed models is briefly outlined for a single parameter value in \autoref{SSHfig2}, where we have, for comparison, shown also the operator obtained from \autoref{SSHnormal} by transforming all elements $-1-\lambda$ into $-1 + \lambda$. All three models exhibit nontrivial domains of observability occurring for non-zero $\gamma$. For non-hermitian operators, it may show rewarding, in addition to studying the spectrum, to examine their $\varepsilon$-pseudospectrum \cite{TrefethenBook,KrejcirikPseudospectra}

\begin{equation}
\sigma_\varepsilon (H) = \left\{ \lambda \in \mathbb{C} \; \big| \; \| (H - \lambda)^{-1} \| \geq \varepsilon^{-1} \right\}
\end{equation}

While the pseudospectrum of a general operator may behave quite arbitrarily, pseudospectra of quasi-hermitian operators admit a simple characterization based on the spectral theorem. Using the decomposition $\Theta = \Omega^\dagger \Omega $, we may write

\begin{equation}
\label{resolventEstimate}
\frac{1}{\rho  ( \lambda, \sigma ( H ))}
\leq
\lVert ( H - \lambda )^{-1}\rVert 
\leq 
\frac{\lVert \Omega \rVert  \lVert \Omega^{-1} \rVert}{\rho  ( \lambda, \sigma ( H ))}
\end{equation}

therefore the $\varepsilon$-pseudospectrum is contained in the $\kappa \varepsilon$-neighborhood of the spectrum, with $\kappa = \lVert \Omega \rVert  \lVert \Omega^{-1} \rVert$. Clearly, the pseudospectrum contains information unavailable from the spectrum itself, with the resolvent norm being pronounced near the eigenvalues, which are likely to complexify by a small perturbation of parameters.

\begin{figure}[h!]
    \centering
    \includegraphics[height=4.7cm,width=7.8cm]{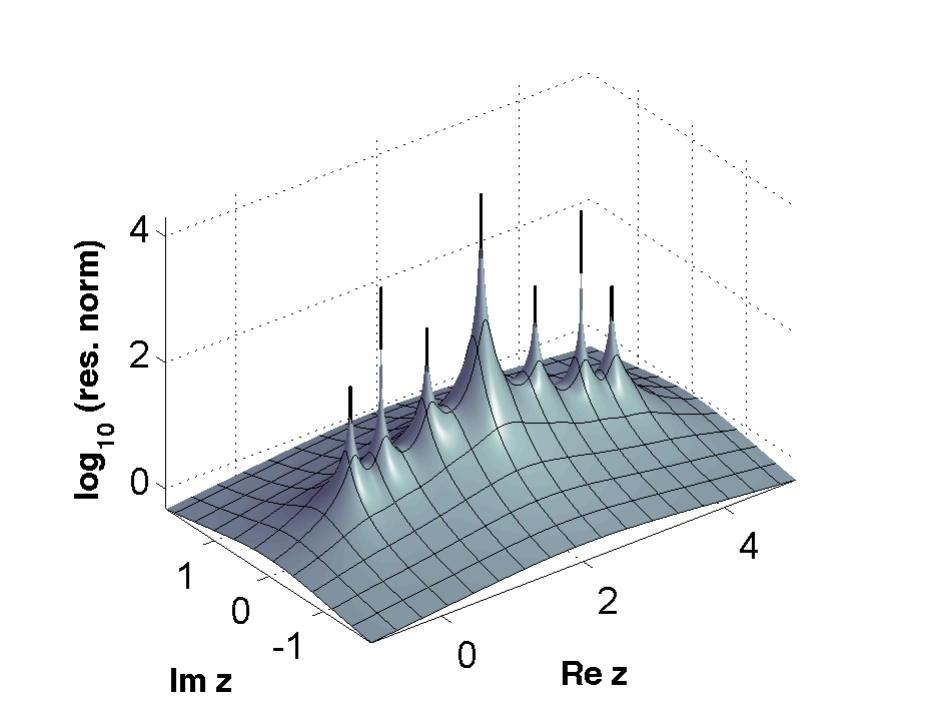} 
    \includegraphics[height=4.7cm,width=7.8cm]{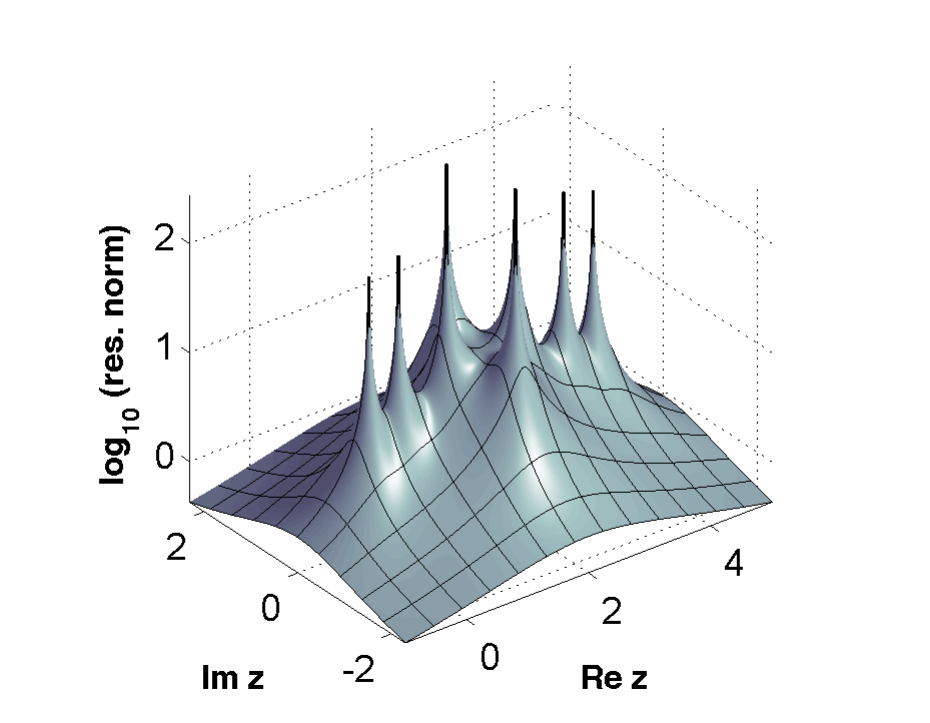} 
    \caption{Pseudospectra of \autoref{discreteKrejcirik} for $n=8$ and $\alpha = 1, \beta=1$, respectively $\alpha = 1, \beta=0.7$}
\end{figure}

\printbibliography

\end{document}